# STATUS OF THE WARM FRONT END OF PIP-II INJECTOR TEST*

A. Shemyakin†, M. Alvarez, R. Andrews, C. Baffes, J.-P. Carneiro, A. Chen, P. F. Derwent, J. P. Edelen, D. Frolov, B. Hanna, L. Prost, A. Saini, G. Saewert, V. Scarpine, V.L.S. Sista[1], J. Steimel, D. Sun, A. Warner, Fermilab, Batavia, IL 60510, USA
[1]also at Bhabha Atomic Research Centre, Mumbai, India

*Abstract*

The Proton Improvement Plan II (PIP-II) at Fermilab [1] is a program of upgrades to the injection complex. At its core is the design and construction of a CW-compatible, pulsed H⁻ SRF linac. To validate the concept of the front-end of such machine, a test accelerator known as PIP-II Injector Test (PIP2IT) is under construction [2]. It includes a 10 mA DC, 30 keV H⁻ ion source, a 2 m-long Low Energy Beam Transport (LEBT), a 2.1 MeV CW RFQ, followed by a Medium Energy Beam Transport (MEBT) that feeds the first of 2 cryomodules increasing the beam energy to about 25 MeV, and a High Energy Beam Transport section (HEBT) that takes the beam to a dump. The ion source, LEBT, RFQ, and initial version of the MEBT have been built, installed, and commissioned. This report presents the overall status of the warm front end.

## PIP2IT WARM FRONT END

PIP2IT realizes the present vision of the PIP-II front end as close as it is possible within building and budget constraints. Correspondingly, its warm portion, when fully assembled (Fig.1), will fulfil all major functions of the PIP-II warm front end.

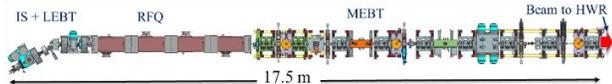

Figure 1: Model of the PIP2IT warm front end (top view).

Presently there are two significant deviations. First, while the PIP-II design employs two ion sources to maximize beam availability, only one is installed at the PIP2IT, though the bending magnet that will be used to switch between the ion sources is installed. Second, the PIP-II MEBT is longer by 3.5 m with three additional triplets and one bunching cavity. The extra length, not included in the PIP2IT MEBT as it is constrained by the existing building, serves mainly two purposes: it improves the protection of the cryomodules from accidental vacuum breaks in the warm section and allows installing a wall to shield the radiation produced by the SRF linac and make the ion source maintenance possible during normal linac operation.

Otherwise, the two front ends are nearly identical. The H⁻ beam is generated in a non-cesiated, filament-driven 30 keV DC ion source [3] capable of up to 15 mA. With the addition of a modulator to the extraction electrode, the source can also provide pulses from 5 μsec to 16 msec in length. Then, the beam enters the LEBT (Fig.2), where it is focused by Solenoid #1, bent by 30º, and prepared for injection into the RFQ with Solenoid #2 and #3. The final pulse length is dictated by a chopper located between the last two solenoids. The chopper can provide 1 μsec -16 msec pulses with a frequency that ranges from single shots to 60 Hz.

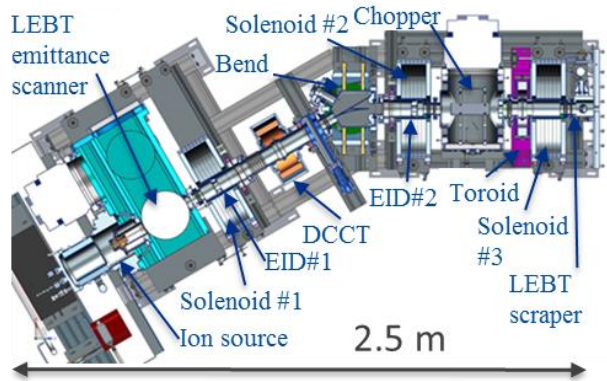

Figure 2: Model of the LEBT (horizontal cross section). EID stands for Electrically Isolated Diaphragm.

The 4-vane, 4 section, 162.5 MHz CW RFQ, designed and constructed at LBNL [4], accelerates the beam from 30 keV to 2.1 MeV and delivers it to the 11 m-long MEBT [5]. Most of the MEBT features are determined by its chopping system that is designed to remove individual bunches according to a pre-determined pattern set to optimize injection into the Booster [1]. The chopping system consists of two travelling-wave kickers separated by 180º transverse phase advance and an absorber at a 90º phase advance from the last kicker. The absorber is followed by a differential pumping section that reduces the flux of gas, released from the bombardment of the absorber with H⁻ ions, into the first cryomodule (Half-Wave Resonator, HWR). Focusing is provided transversely by quadrupoles grouped into two doublets and seven triplets and longitudinally by three bunching cavities, allowing matching of optical functions between the RFQ and HWR. The space between neighbouring triplets is denominated as a section. Each section is 650 mm long flange-to-flange, which was determined based on the space required for the kickers and absorber (500 mm of active length). Each doublet or triplet has a Beam Position Monitor (BPM) button, whose capacitive pickup is bolted to the poles of one of the quadrupoles. Finally, the MEBT includes a scraping system [6] and various diagnostics.

---

## PRESENT BEAM LINE STATUS

The beam line is being assembled in stages, with pace determined by the budget and equipment delivery.

### Ion source and LEBT

The ion source and LEBT have been commissioned first in a straight configuration [7], then with the recent addition of the bend as depicted in Fig. 2. The beam measured at the end of the straight LEBT was within the specifications outlined in [1]. In part, a beam with emittance below 0.13 μm (rms, normalized, for 5 mA) and a current of up to 10 mA DC was demonstrated.

The LEBT employs a scheme with a fully un-neutralized transport in the downstream portion of the beamline [8]. That design combines a good vacuum (~2·10$^{-7}$ Torr) just upstream of the RFQ, low emittance growth, and beam chopping. An important consequence is that the beam characteristics are constant through the pulse [9], which decreases the beam losses in the downstream part of the accelerator and makes measurements performed with short pulses representative of long-pulse or CW operation.

### RFQ

The RFQ was RF commissioned in pulse and CW modes [10]. It works very reliably in pulse mode (typically 10 Hz, 0.1-5 ms) but has two issues in CW. One of them is associated with the power couplers. The RFQ employs two couplers [11] rated for 75 kW CW. However, after delivering ~50 kW in each coupler to the RFQ in CW for several hundred hours, vacuum windows in both couplers sequentially developed vacuum leaks and were replaced by identical spares. While reasons for the failures are not completely understood, running the RFQ in CW is presently suspended until an updated version of the windows is manufactured.

Another issue pertains to the RFQ resonant frequency. Its value is adjusted by regulating the difference in temperature between water cooling the vanes and the body of the RFQ. The middle point of the available tuning frequency range was found shifted by 50 kHz down from the required value of 162.5 MHz, making it uncomfortably close to the regulation boundary. The issue will be corrected by re-machining all 80 RFQ slug tuners.

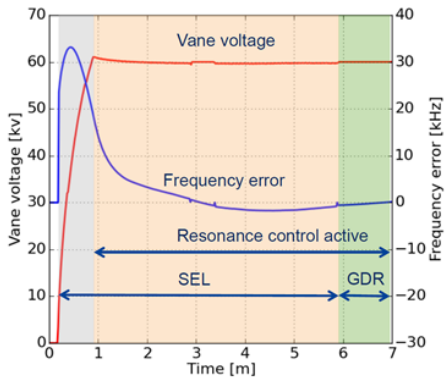

Figure 3: Example of recovery from a 10-seconds RFQ trip.

One of the peculiarities of CW operation is the initial start from a cold state and recovery after RF trips. To address large deviations of the resonant frequency during transitions, an increase in the RFQ power starts with the Low-Level RF system (LLRF) [12] set into self-excited loop (SEL), then the resonance control system brings the RFQ to the nominal frequency by adjusting the vane water temperature [13], and eventually switches to Generator Driven Resonance (GDR) as illustrated by Fig. 3. The cold start and recovery times are well within their corresponding specifications, 30 min and 10 times the RF recovery time, respectively.

### MEBT

The beam accelerated in the RFQ is characterized in a beam line comprising the first MEBT section and a set of diagnostics in several configurations followed by a beam dump. One of the beam line configurations is shown in Fig. 4. The MEBT section contains two quadrupole doublets and a bunching cavity. Most of measurements were made in the short beam pulse mode, 10-20 μs, 10 Hz. At the nominal current of 5 mA and optimum LEBT tuning, the RFQ transmission is 98±2% and the transverse emittance does not exceed the specified value of 0.2 μm (rms, normalized). The ion energy, measured by the time-of-flight method, was found to be 2.11±0.006 MeV. Inconsistencies in the measurements of the bunch length currently prevent asserting a value for the longitudinal emittance. A detailed description of the measurements can be found in Ref. [9].

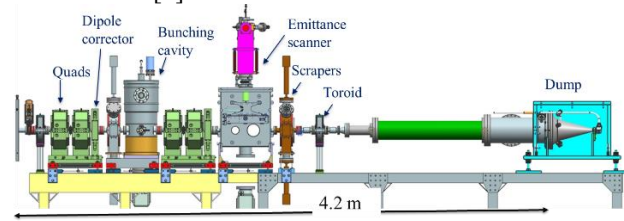

Figure 4: One of the beam line configurations for characterization of the RFQ beam.

The maximum average power delivered to the dump on the time scale of hours was ~5 kW (8 ms x 60 Hz x 5 mA x 2.1 MeV), with another continuous 12-hour run at 2.5 kW. Further power increase attempts were interrupted by a vacuum failure when a bellows was punctured by the beam, in part due to having an incomplete Machine Protection System, and later, by the RFQ couplers failures. The highest power for a 24-hour run was 0.5 kW.

## MEBT DEVELOPMENT

The next step in the development of the PIP2IT warm front end is a longer MEBT for which the main objective is to facilitate characterization of the kickers. Presently being assembled, this configuration adds 4 additional quadrupole triplets, one bunching cavity, two kickers, a set of diagnostics, and a beam dump (Fig. 5). Characterization of the deflection efficiency will be performed with 10 μs pulses, separately for each kicker. Initially every other bunch will be deflected, and the trajectories will be observed by recording signals from the plates of downstream BPMs with fast oscilloscopes.

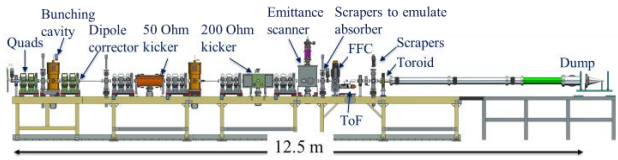

Figure 5: MEBT configuration for kickers characterization. FFC stands for a Fast Faraday Cup, and ToF is a Time-of-Flight monitor.

In addition, the same effect is expected to be clearly visible as a two-hump profile in averaged transverse beam density distributions that can be measured with scrapers or the emittance scanner.

### Kickers

Two kicker prototypes of different designs, termed "50 Ohm" and "200 Ohm" [14] in accordance with their respective impedance, are being installed for testing. To provide a 6σ separation at the absorber between removed and passing bunches, the voltage difference between the opposed electrodes of each kicker should change by 1000 V between those two states.

The 50 Ohm kicker employs plates connected in vacuum by cables (Fig. 6) to slow the propagation of the deflecting signal down to the ions' speed (β= 0.0667).

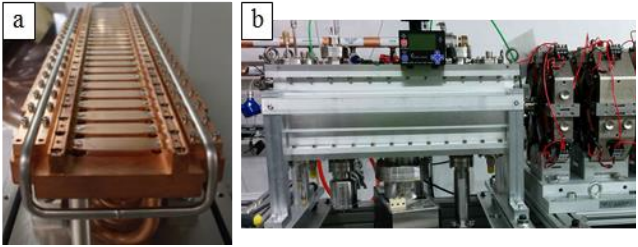

Figure 6: 50 Ohm kicker prototype. **a**- half of the deflecting structure, **b**- the kicker being assembled in the beam line.

In the final scheme, passing and removed bunches are both deflected but in opposite directions with a bipolar signal generated by a commercial wide-band linear CW amplifier. The amplifier's input signal is pre-distorted to account for amplifier imperfections. Such scheme has been demonstrated on a load (with a lower power amplifier) [15]. For the preliminary beam tests, the 50 Ohm kicker will be fed by two narrow-band 81.25 MHz amplifiers so that neighbouring bunches are deflected in opposite directions.

The travelling wave structure of the 200 Ohm kicker is a dual-helix with welded plates facing the beam (Fig. 7a). The higher impedance allows using state-of-the-art fast switches developed at Fermilab [13]. The voltage difference between opposite helices is 1000 V for bunch removal and zero for those passing through. The present version of the switch-driver can deliver an arbitrary pulse pattern that kicks out unnecessary bunches and minimally perturbs the passing ones (Fig. 7b), within 0.6 ms bursts at 20 Hz intervals with a 45 MHz average switching rate. To extend operation into CW, a version with improved cooling of the transistors is under development.

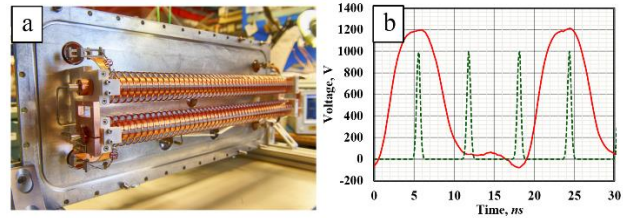

Figure 7: 200 Ohm kicker prototype. a- helices, b- example of a waveform of the difference voltage measured at the helices' outputs (red). Switch-drivers deliver voltages to kick out every third bunch. The green dashed line emulates the Gaussian bunches with nominal width of 0.2 ns rms delivered at 162.5 MHz.

## PLANS

In 2017, the kicker prototypes are planned to be tested and the kicker technology for the PIP-II to be chosen. Furthermore, a full-length MEBT with some elements being still at the prototype level is expected to be assembled and tested. In part, it will include a differential pumping insert. The MEBT absorber rated for 21 kW CW will be designed and manufactured in parallel.

Then operation will be stopped for installation of the cryogenics distribution system and cryomodules. During the shutdown, the final version of the kickers, absorber, fast vacuum protection of cryomodules, and particle-free components of the MEBT will be installed. Beam operation will resume in 2019, with the warm front end initially delivering a pulsed beam into the cryomodules. At that stage, bunch-by-bunch selection is expected to be demonstrated and the absorber test will be performed.

## ACKNOWLEDGMENT

The authors are thankful to the many people who contributed to building PIP2IT and helped in its operation.